\documentstyle[aps,twocolumn,epsfig]{revtex}
\begin{document}
\newcommand{\ve}[1]{\mbox{\boldmath $#1$}}
\twocolumn[\hsize\textwidth\columnwidth\hsize
\csname@twocolumnfalse%
\endcsname

\draft

\title {Low-lying excitations of a trapped rotating 
        Bose-Einstein condensate}
\author{G. M. Kavoulakis$^1$, B. Mottelson$^2$, 
        and S. M. Reimann$^3$} 
\date{\today} 
\address{$^1$Royal Institute of Technology, Lindstedtsv\"agen 24,
         S-10044 Stockholm, Sweden \\
         $^2$NORDITA, Blegdamsvej 17, DK-2100 Copenhagen \O, Denmark \\
         $^3$Division of Mathematical Physics, University of Lund, 22100 
         Lund, Sweden}
\maketitle
					   
\begin{abstract}

We investigate the low-lying excitations of a weakly-interacting, 
harmonically-trapped Bose-Einstein condensed gas under rotation, in 
the limit where the angular mometum $L$ of the system is much less than 
the number of the atoms $N$ in the trap.  We show that in the asymptotic
limit $N \rightarrow \infty$ the excitation energy, measured from the energy
of the lowest state, is given by $27 N_{3}(N_{3}-1) v_0 /68$, where $N_{3}$ 
is the number of octupole excitations and $v_{0}$ is the unit of the 
interaction energy.

\end{abstract}

\vskip0.5pc]
				
The behaviour of a trapped Bose-Einstein condensate under rotation
has attracted much attention in the recent years. References 
\cite{FS,GRPG,Feder,Wilkin,Rokhsar,Ben,BP,WG,LF,KMP,JKMR,JK,NU}
have dealt with this problem theoretically both in the Thomas-Fermi 
limit of strong interactions, as well as in the limit of weak interactions.  
Borrowed from the field of nuclear physics, the terminology ``yrast 
state" refers to the state with the lowest energy for a given angular 
momentum $L$ that the system has.  Of equal importance to the yrast 
state are the low-lying excitations, which determine the thermodynamic 
behaviour of the system, and they are also crucial for the stability 
of the yrast states under external perturbations.

In the present study we consider the hamiltonian $H= H_0 + V$. 
Here
\begin{eqnarray}
    H_0 = \sum_i \left[ - \frac {\hbar^{2}} {2M} {\ve \nabla}_{i}^{2} +
       \frac 1 2 \, M \omega^{2} r_{i}^{2} \right]
\label{h0}
\end{eqnarray}
includes the kinetic energy of the particles and the potential energy
due to the trapping potential, where $M$ is the atom mass, and $\omega$
is the frequency of the harmonic confining potential, which we
assume to be isotropic. Also the interaction between the particles
$V$ is assumed to be short-ranged, 
\begin{eqnarray}
   V = \frac 1 2 U_{0} \sum_{i \neq j} \delta({\bf r}_{i} - {\bf r}_{j}),
\label{int}
\end{eqnarray}
where $U_0 = 4 \pi \hbar^2 a/M$ is the matrix element for atom-atom 
collisions, with $a$ being the scattering length. We consider the limit 
of weak interactions, where the typical interaction energy 
is much less than the typical oscillator quantum of energy,
\begin{eqnarray}
  n U_0 \ll \hbar \omega,
\label{conditionwi}
\end{eqnarray}
with $n$ being the density of atoms.
The above condition allows us to work in the subspace of single-particle
states with no radial excitations,
\begin{equation}
    \Phi_m({\bf r}) = \frac 1 {(m! \pi a_0^3)^{1/2}}
   \left( \frac {\rho}{a_0} \right)^{|m|} e^{i m \phi}
 e^{-(\rho^{2}+z^2)/2 a_0^2}.
\label{phim}
\end{equation}
Here $\rho, z$, and $\phi$ are cylindrical polar coordinates,   
and $a_0 = (\hbar/M \omega)^{1/2}$ is the oscillator length.
These states are
degenerate in the absence of interactions (while states with
radial excitations lie higher by an energy of order $\hbar \omega$).
The whole problem thus reduces to incorporating the interactions
between the atoms, which lift the degeneracy of the states.

One of us has determined in Ref.\,\cite{Ben} the yrast line
of a weakly-interacting Bose-Einstein condensate for an effective 
attraction between the atoms. In the same reference the yrast line has 
also been determined for the case of an effective repulsion 
between the atoms and for $L \ll N$.
In Ref.\,\cite{KMP} we have given a more detailed description of the 
yrast line within the mean-field Gross-Pitaevskii approximation for 
a wide range of values of the ratio $L/N$.  

If one studies the limit $L \ll N$, as shown in Ref.\,\cite{Ben}, in a state 
with a $2^{\lambda}$-pole excitation, 
\begin{eqnarray}
  | \lambda \rangle = | (m=0)^{N-1}, (m=\lambda)^{1} \rangle,
\label{statel}
\end{eqnarray}
where $m$ is the state with angular momentum $m \hbar$, the 
excitation (interaction) energy is 
\begin{eqnarray}
\epsilon_{\lambda} = \langle \lambda | V | \lambda \rangle -  
		     \langle 0 | V | 0 \rangle =
                     - N v_0 \left( 1 - \frac 1 {2^{\lambda - 1}} \right),
\label{elambda}
\end{eqnarray}
where $| 0 \rangle = | (m=0)^N \rangle$ is the ground state, and 
$v_0 = U_0/a_0^3$. The above equation implies that the highest
gain in the interaction energy per unit of angular momentum comes from 
the $\lambda = 2$ or 3 excitations, and therefore the quadrupole and 
octupole excitations are expected to carry the angular momentum for $L 
\ll N$. As a result, in this limit there is a quasi-degeneracy
between the low-lying states, as we discuss below in detail. By
saying quasi-degenerate states, we mean that the energy separation
between them is of order $v_0$, and not of order $N v_0$.  

Bertsch and Papenbrock \cite{BP} have examined the ground state of a
weakly-interacting Bose-Einstein condensate under rotation 
numerically by diagonalizing the hamiltonian $H$ in the subspace of
degenerate states (\ref{phim}) for a given angular momentum. 
In Ref.\,\cite{NU} Nakajima and Ueda have performed similar numerical
calculations in the limit where the angular momentum per particle
$L/N$ is much less than 1, and have found that the quasi-degenerate
states which we mentioned in the previous paragraph lie above the yrast  
by an energy, which in the asymptotic limit $N \rightarrow \infty$ is given 
by $1.59 N_3 (N_3-1) v_0/4$, where $N_3$ is the number of octupole 
excitations.

In this study we give an analytical derivation of this result 
with use of a diagrammatic perturbation-theory approach. The 
starting point of our analysis is the fact that the quadrupole and octupole
$\lambda =2,3$ excitations are dominant for $L \ll N$ \cite{Ben,KMP}, and we 
therefore assume that in this limit the angular momentum is carried by 
$\lambda =2$ and $\lambda =3$ excitations only.  In addition, the 
condensate is dominated by atoms which do not have any
angular momentum.  Our approach therefore consists of 
considering a condensate with $N_0$ atoms in the state 
$m=0$, $N_2$ atoms in the state with $m=2$, and $N_3$ in the state 
with $m=3$, and then treating the other states perturbatively by 
keeping the appropriate diagrams and using perturbation theory to get 
the correction to the energy.  The ``bare" interaction energy with 
particles in the $m=0$, 2 and 3 states is given by (see Fig.\,1 and Table I)
\begin{eqnarray}
{\cal E}^{(0)} = \frac 1 2 N_0 (N_0 - 1)&+&\frac 3 {16} N_2 (N_2 -1)
+ \frac 5 {32} N_3 (N_3 -1) + 
\nonumber \\
&+&\frac 1 2 N_0 N_2 + \frac 1 4 
N_0 N_3 + \frac 5 8 N_2 N_3, 
\label{bare}
\end{eqnarray}
where the interaction energy is measured in units of $v_0$.
The diagrams shown in Fig.\,2 lower the energy, as second-order
perturbation theory implies. For example, for the process 
shown in Fig.\,2(a) where two particles with $m=2$ scatter to a state 
with $m=0$ and $m=4$, and then back to the initial state, the matrix 
element for each vertex is
\begin{eqnarray}
 {\cal M}=\frac {\sqrt 6} {16} \sqrt {N_2 (N_2-1) (N_0+1) (N_4+1)} \,\, v_{0},
\label{mee}
\end{eqnarray}
where $N_m$ is the number of atoms with angular momentum $m \hbar$. 
In Eq.\,(\ref{mee}) we have included the factor $N_4+1$ for clarity,
although as we mentioned above $N_4$ is assumed to be equal to zero.
The result of Eq.\,(\ref{mee}) is most easily derived by writing the 
interaction energy, Eq.\,(\ref{int}), as
\begin{eqnarray}
    V = \frac  1 2 U_0 \sum_{i,j,k,l}
   I_{i,j,k,l}  \, a_i^\dagger a_j^\dagger a_k a_l,
\label{intersq}
\end{eqnarray}
where $a_i$ and $a_i^\dagger$ are annihilation and creation operators
respectively, and
\begin{eqnarray}
   I_{i,j,k,l} &=&
 \int \Phi_i^*({\bf r}) \Phi_j^*({\bf r}) \Phi_k({\bf r})
  \Phi_l({\bf r}) \, d{\bf r}   \phantom{XXX}
\nonumber \\ \phantom{XXX}
  &=& \frac {(i+j)!} {2^{(i+j)}
 \sqrt{i!\, j! \, k! \, l!}} \int |\Phi_0({\bf r})|^4 d{\bf r}
\label{integral}
\end{eqnarray}
when $i+j=k+l$ and zero otherwise. In the case we consider here only
the integrals $I_{0,4,2,2}$ and $I_{4,0,2,2}$ give a non-zero
result reflecting the symmetrization of the bosonic wavefunction. 

  As Eq.\,(\ref{elambda}) implies, the difference between the energy in the 
intermediate and the initial states is $\Delta E= \epsilon_4 + 
\epsilon_0 - 2 \epsilon_2 = -N v_{0}/8$.  
Therefore, according to perturbation theory, the correction to the 
energy is 
\begin{eqnarray}
   \frac {|{\cal M}|^{2}} {\Delta E}  = - \frac 3 {16} N_2 (N_2-1) v_{0}
\label{mnde}
\end{eqnarray}
in the limit $L \ll N$. Table I gives the correction to the energy 
for the processes shown in Fig.\,2. Adding these terms
to ${\cal E}^{(0)}$, we see that the corrected interaction energy 
${\cal E}^{(1)}$ is, in units of $v_0$ and to leading order in 
$N^{-1}$,
\begin{eqnarray}
{\cal E}^{(1)}&=&\frac 1 2 N_0 (N_0 - 1) +  
\frac 5 {34} N_3 (N_3 -1) + \frac 1 2 N_0 N_2 + 
\nonumber \\ &+&\frac 1 4 
N_0 N_3 - \frac 1 4 N_2 N_3 - \frac 3 4 N_3 - \frac 1 2 N_2.
\label{barec}
\end{eqnarray}
To conserve particle number and angular momentum, we have the
following constraints
\begin{eqnarray}
 N_0 + N_2 + N_3 = N \,\,\, {\rm and} \,\,\, 2 N_2 + 3 N_3 =L.
\label{cond}
\end{eqnarray}
Using them, we express ${\cal E}^{(1)}$ in terms of $N, L$, and $N_3$, 
thus finding
\begin{eqnarray}
{\cal E}^{(1)} = \frac 1 4 N (2 N - L -2) + \frac {27} {68} N_3 (N_3 -1).
\label{barecf}
\end{eqnarray}
The number 27/68 coincides with the numerical result 1.59/4 reported in 
Ref.\,\cite{NU}. Here $N_3$ can take all the non-negative integer values 
that are consistent with the constraints of Eq.\,(\ref{cond}). Thus the
number of excited states described by Eqs.\,(\ref{cond}) and (\ref{barecf})
is equal to the integer part of $L/6$.
These statements are exact asymptotically, i.e., 
for $N \rightarrow \infty$, since there are processes which  
couple the quadrupole $m=2$ excitations with the octupole
$m=3$ excitations and they contribute terms of order $1/N$
to the excitation energy. Concerning the yrast state, a consequence 
of Eq.\,(\ref{barecf}) is that its energy is given by the expression
\begin{eqnarray}
  {\cal E}_0 = \frac N 4 (2 N - L -2), 
\label{2syra}
\end{eqnarray}
in agreement with Refs.\,\cite{BP,KMP,JK}. Finally a result of 
Eq.\,(\ref{barecf}) is that the yrast state for $L \ll N$ is
dominated by quadrupole excitations, i.e., to leading order,
\begin{eqnarray}
  \frac {N_2} N = \frac 1 2 \frac L N . 
\label{2syr}
\end{eqnarray} 

In summary we have developed an effective theory which describes the 
ground state and the low-lying excited states of a weakly-interacting
Bose-Einstein condensate under rotation, in the limit where the angular
momentum is much less than the number of atoms. This study demonstrates
that there are low-lying excited states which differ by an energy of 
order $v_0$ (and not $N v_0$), and we have found agreement with a previous
numerical study \cite{NU} of the same problem. We should point out that
for all the values of $L/N$ we have examined numerically, except the
case $L/N \ll 1$, we found that the low-lying excited states are
separated from the yrast state by an energy of order $N v_0$, and in that
respect the limit $L \ll N$ seems to be unique.

\vskip1pc

We would like to thank C. Pethick, A. Jackson and M. Koskinen
for useful discussions. G.M.K. would like to thank the 
Foundation of Research and Technology, Hellas (FORTH) for 
its hospitality. S.M.R. would
like to acknowledge financial support from the ``Bayerische
Staatsministerium f\"ur Wissenschaft, Forschung und Kunst".

\vskip1pc

\begin{figure}
\begin{center}
\epsfig{file=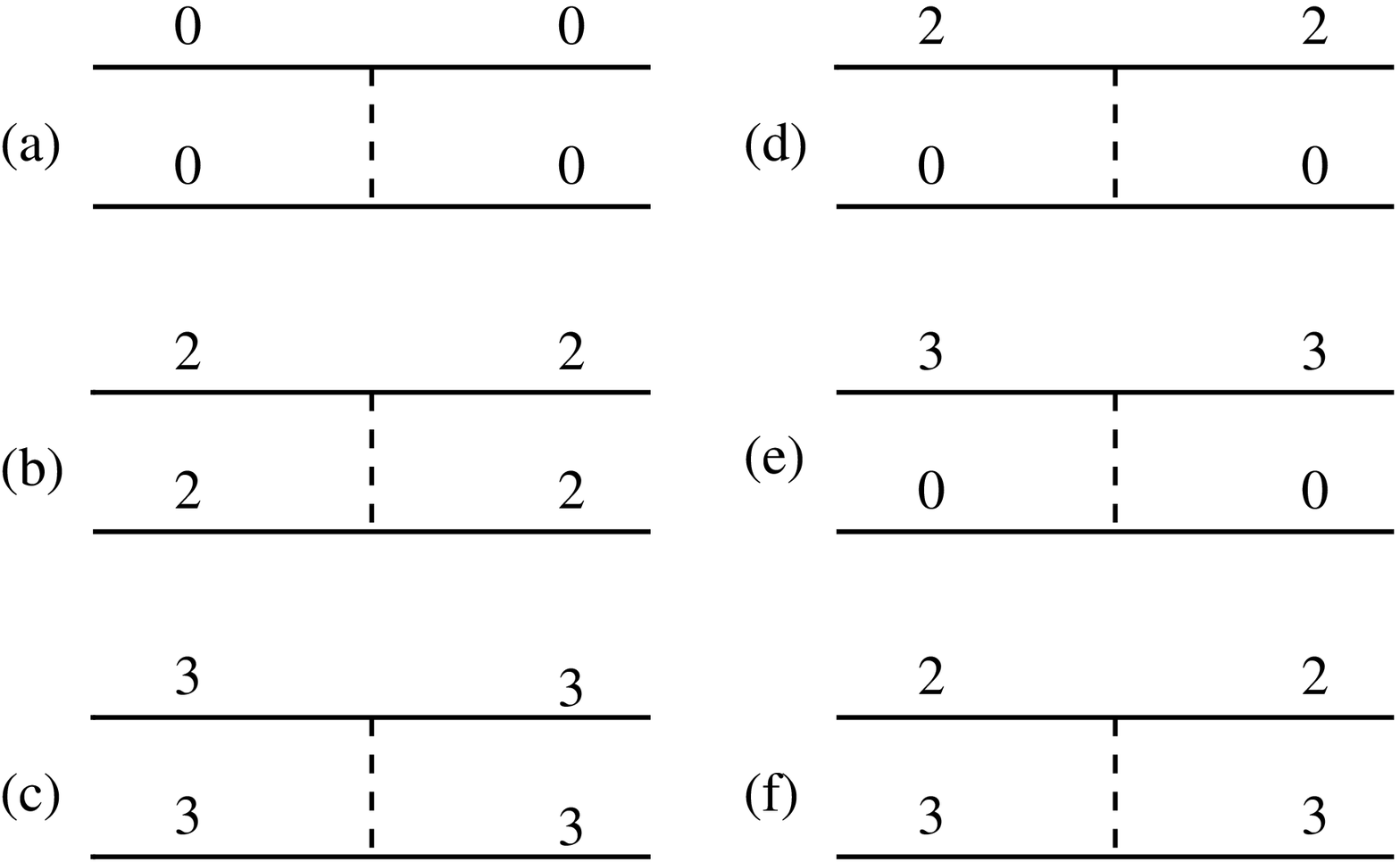,width=8.0cm,height=4.0cm,angle=0}
\vskip1pc
\begin{caption}
{The six diagrams contributing to the bare interaction energy
${\cal E}^{(0)}$, Eq.\,(\ref{bare}). The straight lines denote atoms with
angular momentum given by the numbers written above the lines; dashed lines
denote the interaction.}
\end{caption}
\end{center}
\label{FIG1}
\end{figure}
\begin{figure}
\begin{center}
\epsfig{file=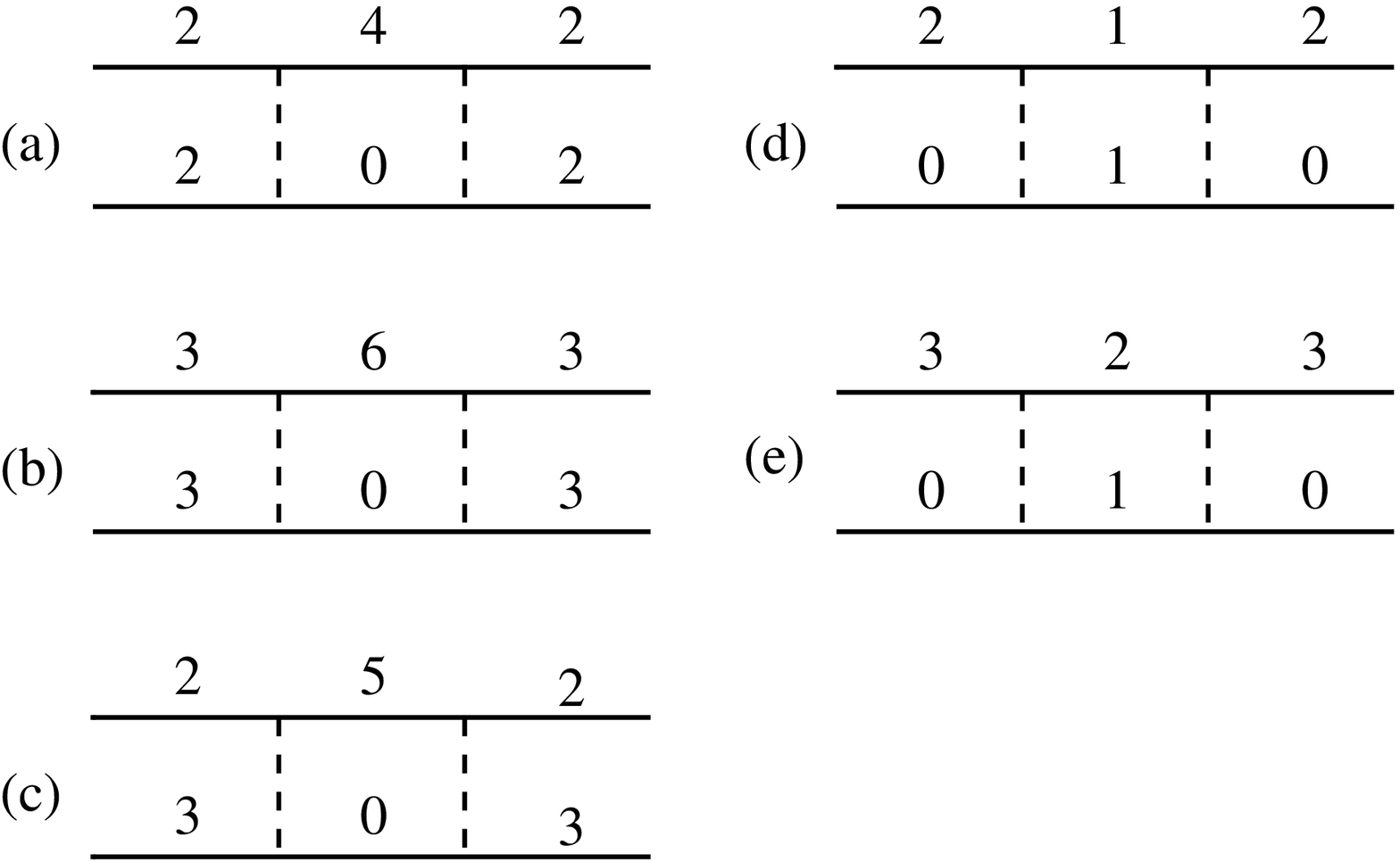,width=8.0cm,height=4.0cm,angle=0}
\vskip1pc
\begin{caption}
{The five (additional) diagrams contributing to the interaction energy
${\cal E}^{(1)}$, Eq.\,(\ref{barecf}).}
\end{caption}
\end{center}
\label{FIG2}
\end{figure}

\begin{table}
\begin{tabular}{||r|l||}
 {\rm Diagram} & {\rm Energy} \\
\hline
1(a) & $N_0(N_0-1)/2$
    \\ \hline
1(b) & $3 N_2(N_2-1)/16$
    \\ \hline
1(c) &$5 N_3(N_3-1)/32$
    \\ \hline
1(d) &$N_0 N_2/2$
    \\ \hline
1(e) &$N_0 N_3/4$
    \\ \hline
1(f) & $5 N_2 N_3/8$
    \\ \hline \hline
2(a)  & $-3 N_2 (N_2-1)/16$
   \\ \hline
2(b)  & $-5 N_3 (N_3-1)/544$
   \\ \hline
2(c) & $-N_2 N_3/8$
   \\ \hline
2(d)& $-N_2/2$
   \\ \hline
2(e)& $-3 N_3 (N_2+1)/4$
\end{tabular}
\vskip0.5pc
\caption{The contribution of the diagrams shown in Figs.\,1 and 2
to the interaction energy (in units of $v_0$).}
\label{table}
\end{table}

\end{document}